\documentclass[prd,a4paper,showpacs,showkeys,preprint,byrevtex]{revtex4}
\usepackage{graphicx}
\usepackage{dcolumn}
\usepackage{amsmath}
\usepackage{bm}

\begin{document}

\title{Gluelump model with transverse constituent gluons}

\author{Fabien \surname{Buisseret}}
\email[E-mail: ]{fabien.buisseret@umh.ac.be}
\thanks{FNRS Research Fellow}
\affiliation{Groupe de   
Physique Nucl\'{e}aire Th\'{e}orique,
Universit\'{e} de Mons-Hainaut,
Acad\'{e}mie universitaire Wallonie-Bruxelles,
Place du Parc 20, B-7000 Mons,  
 Belgium.}
\date{\today}

\begin{abstract}
We show that $C$-odd gluelumps can be successfully described as bound states of a single transverse constituent gluon evolving in the flux-tube-like potential generated by a static color-octet source. The use of a helicity degree of freedom rather than a spin one for the constituent gluon forbids the states that are not observed in lattice QCD. Our model leads to a gluelump mass spectrum in remarkable agreement with the available lattice data provided that an additional parity-splitting mass term is introduced. We argue that such a term is due to instanton-induced interactions in gluelumps.    
\end{abstract}

\maketitle

\section{Introduction}
Quantum chromodynamics (QCD) allows the existence of pure gauge states as glueballs and gluelumps. Glueballs are bound states of the gluonic field alone, while gluelumps, to which this work is devoted, are bound states of the gluonic field plus a static color-octet source. A relative gluelump mass spectrum has been first computed in lattice QCD~\cite{gl2}, while the determination of the absolute masses has been achieved in ref.~\cite{gl1}. Gluelumps were originally a first attempt to model gluon-gluino bound states, allowed by supersymmetric theories~\cite{gl2}. But it has been soon realized that gluelumps are of importance in hadronic physics too. Indeed, if the static source is seen as a pointlike heavy quark-antiquark ($q\bar q$) pair, the gluelump mass can be interpreted as the energy of a static $q\bar q$ pair in the limit where their separation ($r_{q\bar q}$) vanishes. Such an energy has been computed in lattice QCD for $r_{q\bar q}\neq0$~\cite{Juge}, and it indeed appears that the extrapolations of the different energy levels at $r_{q\bar q}=0$ match very well with the gluelump masses~\cite{gl1}. The study of gluelumps is therefore strongly linked to the understanding of heavy hybrid mesons~\cite{Juge,Bramb}, in which there will be an increasing experimental interest with future experiments like BESIII, GLUEX and PANDA. 

Apart from lattice QCD, gluelumps have been studied within different approaches: Bag model~\cite{bag}, Coulomb gauge QCD~\cite{sczg}, pNRQCD~\cite{Bramb}, and potential models~\cite{pot1}. In the framework of potential models, gluelumps are usually seen as a single constituent gluon evolving in the potential generated by the static source. The constituent gluon is generally assumed to have a spin degree of freedom. However, it has been recently shown that the lattice QCD data concerning pure gauge hadrons were rather compatible at a qualitative level with bound states of massless, helicity-1 (transverse), constituent gluons~\cite{boul}. Moreover, several arguments are given in this last reference indicating that a potential model is a relevant approach to study gluonic hadrons. By helicity, we mean that the zero projection of the gluon spin is forbidden. It has already been shown that describing the $C$-even glueballs as bound states of two massless helicity-1 gluons interacting via a flux-tube-inspired potential leads to a good agreement with lattice QCD~\cite{gluh1}. In particular, the use of helicity as intrinsic degree of freedom imposes constraints on the allowed glueball quantum numbers: Light vector glueballs, that are not seen in lattice QCD, are forbidden. This is not the case with spin-1 gluons. 

Motivated by the above discussion, we propose in this paper to reconsider the gluelump spectrum. We first discuss the gluelump quantum states, referred to as gluelump helicity states, in sec.~\ref{hstate}. We then propose a semirelativistic Hamiltonian, based on the flux tube model, in sec.~\ref{potmod}. The $C$-odd gluelump spectrum is numerically computed and discussed in sec.~\ref{conclu}, while a summary of our results is given in sec.~\ref{conclu2}.

\section{Helicity states for gluelumps}\label{hstate}

The lowest-lying gluelumps have a negative charge conjugation and are lighter than the lowest-lying glueballs. We have shown in ref.~\cite{boul} that this fact was compatible with an interpretation of gluelumps as bound states of a single constituent gluon within the potential generated by the static source, while glueballs should be made of at least two constituent gluons. Notice that the static color-octet source has the vacuum quantum numbers $0^{++}$. Our basic assumption is that a constituent gluon is a massless transverse particle. The quantum states describing bound states of such particles can be formulated within the helicity formalism~\cite{jaco}. We refer the reader to this last reference for a detailed description of this formalism. It is enough for our purpose to recall that, within the helicity formalism, there are two families of one-gluon states. Explicitly written in a standard $\left|^{2S+1}L_J\right\rangle$ basis, they read~\cite{boul}
\begin{subequations}\label{hgl}
\begin{eqnarray}
\label{hsgl1}
|A;(J\geq1)^{P-}\rangle &=& \sqrt{\frac{J+1}{2J+1}}\ | ^3 J-1 _J \rangle \nonumber\\
&&+
\sqrt{\frac{J}{2J+1}}\ | ^3 J+1 _J \rangle\nonumber\\
&& \textrm{with} \quad P=(-)^J, \\
\label{hsgl2}
|B;(J\geq 1)^{P-}\rangle &=& -| ^3 J _J \rangle \nonumber\\
&& \textrm{with} \quad P=(-)^{J+1}.
\end{eqnarray}
\end{subequations}
The use of helicity imposes that $J\geq 1$; such a constraint is not present with a spin-1 gluons. The fact that $J=0$ states are forbidden is in agreement with lattice QCD since no light $0^{P-}$ gluelump is observed~\cite{gl2}. Moreover, the lightest gluelumps are $1^{P-}$ ones as expected from eqs.~(\ref{hgl}). Notice that the helicity formalism also reduces the number of possible gluelump states, and thus the arbitrary of the model. For example, $|A;1^{--}\rangle$ is the unique $1^{--}$ gluelump helicity state, while both $| ^3 S _1 \rangle$ and $| ^3 D _1 \rangle$ would be allowed with spin-1 gluons.         

Thanks to the above decompositions, simple computations show that the matrix elements of the different operators usually appearing in potential models (square orbital angular momentum, spin-orbit, tensor, etc.) have the same values for both families. One has indeed $\left\langle \bm L^{\, 2} \right\rangle=J(J+1)$, $\left\langle \bm S^2\right\rangle=2$, $\left\langle \bm L\cdot\bm S\right\rangle=-1$, and $\left\langle \bm S^2-3(\bm S\cdot\hat{\bm r})^2\right\rangle=-1$, where the average values are computed with either the state~(\ref{hsgl1}) or the state~(\ref{hsgl2}).

Gluelumps with a positive charge conjugation cannot be one-gluon states: At least two gluons are needed~\cite{sczg,boul}. This is coherent with the fact that the currently known $C$-even gluelumps ($0^{++}$ and $1^{-+}$) are well heavier than the $C$-odd ones. The corresponding helicity states can be obtained by coupling the two constituent gluons to a given total spin $j$ and parity $\pi$, and by recoupling these two gluons, seen as a massive particle of spin $j$ and parity $\pi$, to the static source following a general procedure described in ref.~\cite{wick3}. General two-gluon states are extensively studied in ref.~\cite{gluh1}. Here it is sufficient to recall that the lightest totally symmetric two-gluon configuration is reached for $j^\pi=0^+$. The two-gluon cluster and the static source, being both massive objects, can be directly coupled  in the usual $\left|^{2S+1}L_J\right\rangle$ basis. The two lightest states should be mainly the $\left|^1S_0\right\rangle$ and $\left|^1P_1 \right\rangle$ ones, which have the quantum numbers $0^{++}$ and $1^{-+}$ respectively, in agreement with lattice QCD. Notice that in the following we focus on $C$-odd gluelumps only. 

\section{The model}\label{potmod}

At the dominant order, the flux tube Hamiltonian describing a gluelump made of one massless constituent gluon reads
\begin{equation}\label{gluelu}
	H_0=\sqrt{\bm p^{\, 2}}+ar-3\frac{\alpha_s}{r},
\end{equation}
where $\bm p$ is the momentum of the constituent gluon and where $\bm r$ is the separation between the constituent gluon and the static source. Notice that $r=|\bm r\,|$. This Hamiltonian is spin-independent but the gluon helicity is already included at the level of the gluelump helicity states~(\ref{hgl}). The formalism is not explicitly covariant, but the semirelativistic kinetic term $\sqrt{\vec p\,^2}$ is relevant to describe the dynamics of a massless particle. $a$ is the energy density of the straight flux tube linking the constituent gluon to the static source. Assuming the Casimir scaling hypothesis, one can write $a=(9/4)\sigma$, with $\sigma$ the energy density of a fundamental flux tube, that is the flux tube in a meson. The flux tube stands for the nonperturbative part of the QCD interactions and generates the confinement. It is a dynamical object that carries both energy and angular momentum but, at the lowest order, it reduces to a linearly rising confining potential. The flux tube clearly dominates the long range interactions. Nevertheless, other mechanisms come into play at short distances, i.e. one gluon exchange processes. The Coulomb part, where $\alpha_s$ is the strong coupling constant, is the dominant part of the one gluon exchange potential between the constituent gluon and the static source~\cite{oge}.  

Lattice QCD mass spectra are generally expressed in units of a fundamental length scale $r_0$ that can be linked to $\sigma$ by the relation $\sigma=1/r^2_0$. Most of the final lattice error bars actually comes from the determination of this length scale $r_0$, and it is thus worth comparing our model with the dimensionless lattice QCD spectrum $r_0\, M_{\rm{Lat}}$. To this aim, it is convenient to work with the new coordinates $\bm q=r_0\, \bm p$ and $\bm x=\bm r/r_0$. Then, Hamiltonian~(\ref{gluelu}) becomes 
\begin{equation}\label{gl2}
	r_0\, H_0=\sqrt{\bm q\,^2}+\frac{9}{4}x-3\frac{\alpha_s}{x}.
\end{equation}
This rescaling eliminates a free parameter in our model: Only $\alpha_s$ has now to be fitted on the data. Before making explicit numerical computations, it should be written that 
\begin{equation}\label{matelem}
	\bm q^2=q^2_x+\frac{\left\langle \bm L^2\right\rangle}{x^2}=q^2_x+\frac{J(J+1)}{x^2}.
\end{equation}
This particular value of the square orbital angular momentum is a consequence of the helicity states~(\ref{hgl}), as we already mentioned in the previous section. 

The Hamiltonian~(\ref{gl2}) only depends on $J$ through eq.~(\ref{matelem}). It means that the gluelumps of quantum numbers $J^{\pm-}$ are degenerate if only $H_0$ is used to compute the gluelump mass spectrum, in obvious contradiction with lattice QCD. Adding the usual spin-dependent corrections of Fermi-Breit type would not lift this degeneracy since, as shown in sec.~\ref{hstate}, the matrix elements of the spin-spin, spin-orbit, and tensor operators are identical for both gluelump helicity states. This leads us to introduce a phenomenological mass term of the form 
\begin{equation}\label{inst}
r_0\, {\rm \Delta} M=(-)^J\, r_0\, \gamma\, P,	
\end{equation}
splitting the gluelumps with opposite parity but with the same total spin $J$. \textit{A priori}, $\gamma$ could be a complicated function of $J$, $x$, \textit{etc}. Nevertheless, we assume here that $\gamma$ is constant and positive. The meaning of this extra term will be discussed in the next section. Notice that the need for an interaction term going beyond the standard pair interactions in gluelumps has also been pointed out in the Coulomb gauge study of ref.~\cite{sczg}.

\section{Results and discussion}\label{conclu}

The mass spectrum of Hamiltonian~(\ref{gl2}), denoted as $r_0\, M_0$, can be accurately computed by using the Lagrange mesh method~\cite{sem01}. Adding the mass term~(\ref{inst}), one can finally obtain the total gluelump mass
\begin{subequations}\label{eqsol}
\begin{equation}\label{maineq}
	r_0\, M=r_0\, M_0+r_0{\rm \Delta} M,
\end{equation}
where $r_0\, M_0$ is the eigenvalue of Hamiltonian~(\ref{gl2}).

As it can be seen in table~\ref{tab}, a remarkable agreement between our model and the available lattice QCD data is reached provided that the following values are used for the parameters
\begin{equation}\label{params}
	\alpha_s=0.38,\quad r_0\, \gamma=0.47.
\end{equation}
\end{subequations}
Note that $\alpha_s\approx0.4$ is a standard value in light hadron physics. In particular, the light meson~\cite{buis} and glueball~\cite{gluh1} mass spectra can be satisfactory reproduced within the flux tube model by using such a value for the strong coupling constant. Notice that the results of table~\ref{tab} can be converted in physical units with the typical conversion factor $r_0^{-1}=440$~MeV. One has then $\gamma\approx0.2$~GeV.

\begin{table}[t]
	\centering
	\caption{Gluelump spectrum computed from eqs.~(\ref{eqsol}) compared with the lattice QCD data of ref.~\cite{gl1}.}
		\begin{tabular}{c|c|c}
		$J^{PC}$ & $r_0\, M_{\rm{Lat}}$~\cite{gl1} & $r_0 M$	\\
		\hline
		$1^{+-}$ & 2.25(39) & 2.25 \\
		$1^{--}$ & 3.18(41) & 3.19 \\
		$2^{--}$ & 3.69(42) & 3.70 \\
		$2^{+-}$ & 4.72(48) & 4.64 \\
		$3^{+-}$ & 4.72(45) & 4.70 \\
		$4^{--}$ & 5.41(46) & 5.52 \\
		$3^{--}$ &          & 5.64
		\end{tabular}
	\label{tab}
\end{table}
\begin{table}[t]
	\centering
	\caption{Mass gaps~(\ref{massg}) computed from eqs.~(\ref{eqsol}) and compared with the lattice QCD data of ref.~\cite{gl2}.}
		\begin{tabular}{c|c|c}
		$J^{PC}$ & $r_0\, M^R_{\rm{Lat}}$~\cite{gl2} & $r_0 M^R$	\\
		\hline
		$1^{--}$ & 0.933(18) & 0.94 \\
		$2^{--}$ & 1.438(25) & 1.45 \\
		$2^{+-}$ & 2.467(92) & 2.39 \\
		$3^{+-}$ & 2.468(60) & 2.45 \\
		\end{tabular}
	\label{tab2}
\end{table}
Without the phenomenological term~(\ref{inst}), the masses of gluelumps with the same total spin but opposite parity would be equal, in disagreement with lattice QCD. Once it is introduced, it first reproduces the correct ordering of the $1^{\pm-}$ and $2^{\mp-}$ states. Second, it lowers the $3^{+-}$ mass so that it is nearly degenerate with the $2^{+-}$, as observed in ref.~\cite{gl1}. Finally, it causes the $4^{--}$ gluelump to be lighter than the $3^{--}$ one. This is not an intuitive result since Hamiltonian~(\ref{gl2}) alone states that gluelumps with higher $J$ should be always heavier than states with lower $J$. 

The results of our model can also be compared with the relative gluelump mass spectrum that has been computed in ref.~\cite{gl2}. In this last reference, the mass of the $1^{+-}$ gluelump ($M_{1^{+-}}$) is not known, and only the mass gaps between the various gluelump states and the $1^{+-}$ one can be computed. But, the accuracy of the results is better than in ref.~\cite{gl1}. We denote
\begin{equation}\label{massg}
	r_0\, M^R=r_0\, M-r_0\, M_{1^{+-}}
\end{equation}
 the mass gaps. As it can be observed in table~\ref{tab2}, our predicted values compare very favorably with lattice QCD. In particular, the mass gaps concerning the $2^{--}$ and $3^{+-}$ gluelumps are independent of the term~(\ref{inst}). The fact that they are compatible with the lattice results is a good check of the relevance of Hamiltonian~(\ref{gl2}) alone.

Although leading to nice results, the mass term~(\ref{inst}) is introduced \textit{ad hoc}. Obviously, its origin cannot be linked to some short-range correction of one gluon exchange type since all the operators appearing in such interactions only depend on $J$, not on $P$. One could think about higher-order corrections, with operators such as $\bm L^4$ for example, but these corrections are not proportional to $P$ in general. Consequently, this term should be nonperturbative, and instanton-induced forces are a relevant candidate. Indeed, as a consequence of the self-duality of the instanton's field strength, it has been argued that the instanton contribution is generally equal in magnitude but opposite in sign for hadrons with the same total spin and charge conjugation, but with opposite parity~\cite{Kochelev:2005vd}, just as the mass term~(\ref{inst}) is. Although instanton-induced interactions are well known for mesons, it is not the case for gluonic hadrons. From refs.~\cite{inst}, it can be deduced that instantons induce a mass term of the form $- {\cal I}\, P\, \delta_{J,0}$ for $C$-even glueballs. Such a term has been successfully used in ref.~\cite{gluh1}, which is clearly a two-gluon generalization of the present work. We are thus led to conjecture that our phenomenological term~(\ref{inst}) originates in instanton-induced forces in gluelumps. To our knowledge, no study of these interactions has been performed yet and this is a task out of the scope of this paper. We hope that such results will be available in the future in order to check if the form $(-)^J\, \gamma\, P$ eventually emerges as an effective instanton mass term in gluelumps. 

The $(-)^J$ factor in eq.~(\ref{inst}) could seem rather surprising. It is actually such that, for a given $J$, the helicity state (\ref{hsgl2}) is always lighter than the state~(\ref{hsgl1}). The $\left|B\right\rangle$ family, the lightest one, corresponds to the gluelump states of natural parity, that is $\eta_g\, (-1)^J$, $\eta_g$ being the intrinsic parity of a gluon. We already pointed out in ref.~\cite{boul} that the gluonic states appearing in lattice QCD (glueballs and gluelumps) are always such that those with natural parity are lighter. Our mass term~(\ref{inst}), as well as the currently known gluelump spectrum, is in agreement with this fact.

\section{Conclusion}\label{conclu2}

In summary, we have described the $C$-odd gluelumps as bound states of one transverse gluon within the potential generated by a static color-octet source. The gluelump quantum states are computed thanks to the helicity formalism, and the Hamiltonian corresponds to the dominant order of the flux tube model with one-gluon exchange potential. A very good agreement with lattice QCD is reached provided that an extra term $(-)^J\, \gamma\, P$ is introduced. Although further research in that domain is needed, we expect that this term is intrinsically nonperturbative and comes from instanton-induced interactions in gluelumps. 

\acknowledgments
The author thanks Claude Semay for valuable discussions about the present work.

\end{document}